# *Transparent and Traceable Food Supply Chain Management*

Narayan Subramanian
*School of Computer Science and Technology*
*Vellore Institute of Technology, Chennai,India*
narayan.subramanian2020@vitstudent.ac.in

Atharva Joshi
*School of Computer Science and Technology*
*Vellore Institute of Technology, Chennai,India*
atharvavivek.joshii2020@vitstudent.ac.in

Daksh Bagga
*School of Computer Science and Technology*
*Vellore Institute of Technology, Chennai,India*
daksh.bagga2020@vitstudent.ac.in

*Abstract*— The food supply chain has a number of challenges, including a lack of transparency and disengagement among stakeholders. By providing a transparent and traceable digital ledger of transactions and movements for all supply chain actors, blockchain technology can provide a resolution to these problems. We propose a blockchain-based system for tracking a product's full path, from its raw components to the finished item in the store. Many advantages of the offered system include improved quality assessment, increased product transparency and traceability, and sophisticated fraud detection capabilities. By reinventing the way transactions are carried out and enabling stakeholders to obtain a complete record of each product's journey, the system has the potential to completely alter the food supply chain. Also, by minimising inefficiencies, waste, and fraudulent activities that have a negative influence on the supply chain, the deployment of this system can remove limits imposed by the current supply chain. Overall, the suggested blockchain-based system has the potential to significantly increase the efficiency, transparency, and traceability of the food supply chain.

## I. Introduction

The food supply chain is an intricate and multifaceted network with many different stakeholders, all of whom are essential to making sure that food products get to consumers. Despite recent considerable technical improvements, there is still a lack of transparency in the food supply chain at the manufacturing and distribution level. The sector faces many difficulties as a result of this lack of transparency, including inefficiencies, waste, and fraudulent actions that may compromise the integrity of the entire supply chain. Lack of comprehensive product traceability to the original source is one of the biggest problems facing the present food supply chain.

Keeping track of the origins, steps taken, and people involved in the handling of food goods when they are transferred from one stakeholder to another can be difficult. As a result, it can be challenging to guarantee product quality and safety because items may be the target of adulteration, mislabeling, or other fraudulent actions.

We therefore suggest a blockchain-based system that offers a transparent and verifiable supply chain network as a solution to these problems. All stakeholders, including customers, may access and examine a transparent and unchangeable digital ledger of transactions and movement thanks to blockchain technology. With the help of the suggested system, stakeholders will be able to follow a specific product all the way from the farm where its raw materials, such tomatoes, are grown to the retailer store where it is processed and put up for sale.

By providing a number of advantages, the suggested blockchain-based solution will transform the food supply chain. First off, the technology will enhance how products are evaluated for quality. The harvest includes several quality reports, which will be kept in a blockchain network. Processors might utilise these data as a criterion to judge the calibre of raw materials. Second, the system will improve the items' transparency and traceability. The processor adds the relevant reports (along with a timestamp) to a blockchain network after receiving the raw materials. This will make it possible for all parties involved to examine all records, including quality, processor, and retailer reports, from the store to the farmer before buying a product. The suggested method will also improve fraud prevention capabilities. As the system is completely transparent and each step includes a timestamp, any fraud or forgery (even hoarding) can be tracked. Stakeholders may confirm product authenticity and guarantee that goods are not tampered with or compromised in any manner by utilizing blockchain technology

The remainder of the paper is organized in the following Sections. Section II describes the existing architecture followed by Section III which describes the infrastructure. Section IV contains information about blockchain followed by Section V which tells us more about Federated Learning. Section VI describes the proposed architecture followed by the concluding remarks of the study and the future works in Sections VII and VIII.

## II. Abbreviations and Acronyms

1. IoT – Internet of Things
2. MLAV - Multi-Layer Aggregate Verification
3. SCOR - Supply Chain Operations Reference
4. AHP - Analytic Hierarchy Process
5. FCA - Fuzzy Comprehensive Analysis

## III. Existing Architecture

### A. Related Works

The utilization of blockchain technology can address the challenges faced by agricultural producers, particularly in relation to supply chain management. Hegde et. al [1] emphasize the need for a reliable database to transfer knowledge in the agricultural industry. The use of blockchain technology can mitigate the spread of misinformation by providing a trustworthy and incorruptible data ledger. A model incorporating blockchain technology into the agricultural supply chain as a transparent and dependable transaction mechanism is proposed. Smart contracts can be used to ensure that all parties agree and deliver their parts, without marginalizing any one tier. The use of blockchain technology in the Indian agricultural supply chain can lead to increased efficiency, decreased waste, and an overall improvement in the industry.

Tiaobin et.al [2] proposed three specific applications of the IOT in fresh agricultural product supply chain management: monitoring of fresh agricultural products, strict quality control to ensure food safety, and the creation of a management information system based on IOT to increase supply chain integration. The EPC is used as the foundation of IOT operations to provide unique identities for physical objects, which can improve monitoring levels for commodity production, distribution, warehousing, and sales. The IOT has the potential to completely change supply chain processes and management methods, providing new opportunities for the development of supply chain management in enterprises. Efficient fresh agricultural product supply chain management is key to improving the competitiveness of fresh produce enterprises.

Shen et. al [3] examines the importance of third-party certification agencies in supplier selection for supply chain management, particularly in situations with incomplete market information. Using the signalling game method, a mathematical model is established to analyse the dynamics of the market and discuss the equilibrium. The study finds that in a separating equilibrium, the signal conveyed by the supplier represents their true type, which allows for the efficient selection of certified suppliers by the buyer, while speculative suppliers are discouraged from participating due to prohibitive costs. The most ideal and efficient equilibrium is the separating equilibrium, where the supplier is subject to "telling the truth" And the market's performance is maintained. The study highlights the significance of third-party certification agencies in ensuring the credibility of the supplier selection process.

Sihuan et al [4] focused on the vulnerability of the agricultural product supply chain in the Eastern Area of Hunan Province and proposes a risk management approach based on the AHP-FCS method. The methodology consists of two parts: the AHP and the method FCS. The AHP-FCS model provides a more reliable and efficient approach for identifying and assessing agricultural product supply chain risks. The conclusion suggests that simple risk assessment is not enough and that a scientific and practical risk tracking system is necessary for effective agricultural product supply chain management.

Wang et al [5] presented decision models to analyze the impact of supply chain coordination on the order quantity and ordering cycle for deteriorating goods with stock-dependent demand rates. Two scenarios, decentralized and centralized supply chains, are considered. In the decentralized supply chain, each entity maximizes its own profit functions, while in the centralized supply chain, the order quantity and replenishment cycle are determined to maximize the overall profit incurred by both the retailer and the manufacturer. The efficacy of the proposed models is demonstrated through a numerical study, and sensitivity analysis is conducted to investigate the impact of various model parameters on the supply chain profit increase percentages generated by supply chain coordination. The study shows that a centralized policy is always more efficient than a decentralized policy, in terms of supply chain profit.

Yuniaristanto et al [6] aims to address the issues faced by Vietnamese coffee's competitiveness by proposing a standard coffee supply chain model using the SCOR model. The model is used to investigate external and internal issues that reduce the efficiency of the coffee supply chain in Kontum province, Vietnam, and to measure performance in the coffee supply chain using the SCOR model. The case study methodology was used to extend the SCOR model and validate the suitability of developed models. The study found that the overall coffee supply chain performance is 68.28, with an average category, and most processes have a low performance value. The proposed model can assist supply chain members, particularly coffee companies in other countries with similar coffee supply chains to Vietnam, map and evaluate the supply chain's success.

Harding et al [7] presented a MLAV solution for IoT Blockchain devices to improve supply chain management in Agriculture 4.0. The existing blockchain solutions are inadequate as they only serve large-scale production suppliers and do not facilitate smallholders' participation in the agricultural blockchain. The proposed solution employs a multi-layer architecture to allow smallholders' participation and reduce costs. The methodology involves the use of periodic firmware updates and an aggregate verification method to efficiently verify many signatures at once. The proposed solution significantly reduces network traffic from IoT devices on the blockchain network and shifts computing overhead to aggregator nodes. This framework reduces the workload of all participants and speeds up the agricultural industry's financial processes.

Arora et al [8] proposed a new blockchain implementation for tracking the carbon footprint of food production and transportation stages to mitigate the effects of increased food demand on the environment and climate. The proposed system uses cluster-based record keeping to track the carbon footprint of food processing facilities and transportation parties while protecting their privacy. The system uses a Raft-like consensus algorithm to arbitrate decisions on leader election within a cluster, node addition, and block updates. The carbon footprint chain is divided into six clusters, each representing a stage of the food life cycle, and the blockchain nodes are the facilities within each life cycle stage. The proposed system enables lightweight distributed-record keeping for tracking carbon footprint in food transportation, which occurs when food is transported from one stage to the next in the food life cycle.

Malik et al [9] highlighted the importance of agri-food supply chain traceability and proposes a generic framework for a traceability model that includes blockchain as a key component. The framework includes building blocks from the data, storage, application, and blockchain layers, and the paper discusses the challenges and potential benefits of using blockchain to improve supply chain resilience. The methodology takes a holistic design approach and shows which physical entities in the supply chain the technical components must be linked to. The paper emphasizes the need for blockchain-assisted agri-food supply chain traceability systems to guarantee data authenticity and immutability and to address challenges such as compromised food safety, data modification fraud, counterfeiting, and food waste.

*B. The Existing System*

In India iinefficiency and losses in the supply chain are observed and are caused by a fragmented supply chain. India's agricultural productivity has increased by 40% to 500% in the last 40 years. Food availability, on the other hand, still remains a major concern due to loopholes in supply chain. Quality control is a significant challenge due to the massive size of the supply chain.

The absence of technology has a significant impact on the supply chain lead time in India. The lack of organised logistics impedes the transportation of produce from farms to consumers. In India, some compliance is lost in the traditional route of buying from traders and wholesalers. People frequently misunderstand the relationship between sustainability and traceability. The ability to track materials from the beginning of the supply chain to the customer who purchases a product is referred to as traceability. Traceability provides visibility across the value chain on inputs and processes, as well as the source information for its origins and sustainability certifications. Traceability using a blockchain based system is still in a nascent stage and has still not been addressed properly.

## IV. Proposed Methodology

The proposed implementation process for blockchain technology in agricultural supply chain management entails a number of steps. First, the farmer and the buyer will agree on a smart contract defining the specifics of the transaction. After that, the commodities will be transported from one place to another with an RFID tag connected to the produce consignment.

As a product reaches a warehouse, the RFID tag will be read, updating the position, allowing for real-time tracking of the product as it moves through the supply chain. The blockchain will be used to securely store the location information as well as other quality indicators like temperature and humidity. The RFID tag will continue to update the location and quality data on the blockchain as the product continues to move through the supply chain. The smart contract will be satisfied and the transaction will be finished once the item arrives at its destination.

The proposed solution will revolutionize the entire supply chain by transforming the current supply chain by following implementations :-

• A lot's quality reports (part of the harvest) are stored in a blockchain network. These reports can be used by processors as a metric to assess the quality of raw materials.

• After receiving raw materials, the processor adds the respective reports (along with a timestamp) to a blockchain network.

• Before purchasing a product, the customer can review all the reports from retailer to farmer (quality reports -> processor report -> retailer report).

• The system is completely open. Because the timestamp is recorded at each step, any forgery or other fraudulent activity (hoarding) can be traced.

Overall, the suggested methodology will offer a safe and open system for following agricultural products along the supply chain, making sure they are of excellent quality and get to where they are supposed to. Farmers and buyers may increase their trust in the supply chain and make sure that the products are handled appropriately throughout the entire process by leveraging blockchain technology and smart contracts.

*A. Novelty in concept*

The use of blockchain technology in agriculture supply chain management is a novel approach that brings several unique benefits to the industry. Here are some of the key ways in which blockchain technology is novel in the context of agriculture supply chain management:

1. Trust and transparency: Blockchain technology provides a highly secure and transparent system for tracking the movement of agricultural products from the farm to the consumer.

2. Smart contracts: The use of smart contracts in agriculture supply chain management is a novel approach as by automating the execution of contracts, smart contracts can reduce the need for intermediaries and increase the speed of transactions and overall improve the efficiency of the supply chain.

3. Provenance tracking: Help to ensure that products are authentic and have been produced in compliance with relevant standards and regulations. It can help to reduce the risk of fraud and increase the trustworthiness of the supply chain.

4. Supply chain optimization: Reduce inefficiencies and minimize waste. By providing real-time data on the movement and quality of products, it can help to optimize logistics and reduce the need for manual interventions.

*B. Feasibility*

To make this idea a feasible solution, it requires immense government support and a heavy capital investment to set up the entire ecosystem. From a technical viewpoint, the project is feasible. But some obstacles which we can face in this are the blockchain might get overloaded with data because as the package travels, at every point the data is being fed into the blockchain. From an economical viewpoint, this is going to be expensive to set up but considering that government will invest and support in this project, it is an idea which can be implemented. Another problem which we can encounter is that it is hard to convince people at first because blockchain is very new. India has come a long way in the last decade in terms of food traceability, thanks to private and public sector initiatives.

Food traceability, once implemented across supply chains, will have the following effects on our supply chains:
• Meet consumer demand for transparency in food production.
• Improve your ability to recognise, respond to, and even prevent food safety issues.
• Help to optimise the supply chain and reduce food waste
• Validate sourcing claims to achieve sustainability goals.

## V. REQUIREMENTS AND OUTPUTS

- Hardware

    Rfid Tag
    Rfid Scanner
    Arduino
    Servo motor
    Jumper cables and LEDs

- Software

    Ethereum
    MetaMask
    Truffle
    NodeJS
    Infura.

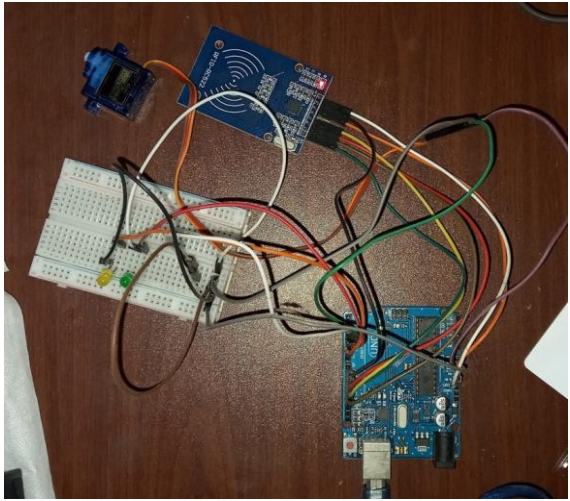

**Fig 2**
**Circuit connections**

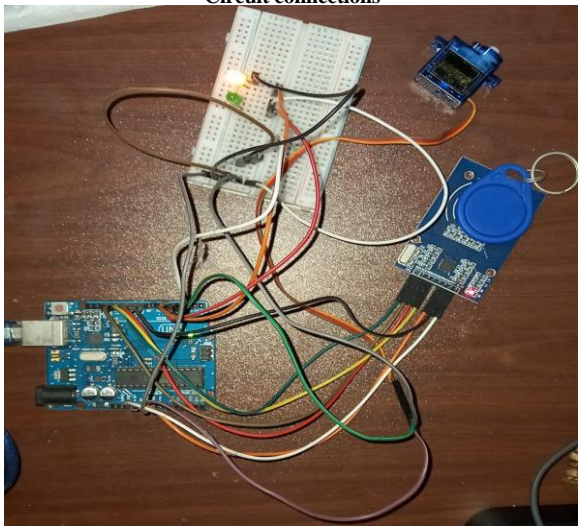

**Fig 3**
**Red Led blinking when wrong RFID tag is scanned**

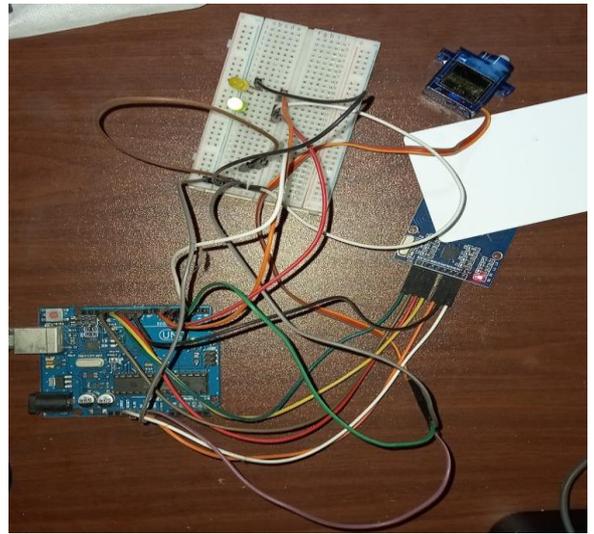

**Fig 4**
**Green Led blinking when right RFID tag is scanned**

```
Put your card to the reader...

UID tag : F3 82 95 2E
Message : Authorized access
```

**Fig 5 (a)**
**Authorized access**

```
UID tag : 70 DF 68 21
Message : Access denied
```

**. Fig 5 (b)**
**Un-authorized access**

## VI. CONCLUSION AND INFERNECES

The food supply chain can be revolutionized by a blockchain-based system by increasing efficiency, transparency, and traceability, which will have a favorable effect on public health, the environment, and the economy. Each stage of the supply chain may be tracked by the system, allowing early detection of potential quality problems, preventing foodborne diseases, decreasing food waste, and informing consumers about the country of origin of their food.

Additionally, the use of a blockchain-based system can aid in the prevention of fraudulent activities like forging, mislabeling, and tampering by spotting any discrepancies in the supply chain and enabling quick remedial action. This project has a bright future ahead of it, with a wide range of opportunities for growth and collaboration with other new technologies, governmental organizations, and international marketplaces that will result in a more ethical and sustainable global supply chain. Ultimately, a blockchain-based system for the food supply chain might be advantageous to all parties and help it become more efficient, transparent, and sustainable.

## VII. FUTURE WORKS

The future scope of a blockchain-based system for the food supply chain is vast and promising. Here are some potential areas where this project can be expanded and developed further:

- Integration with other technologies.
- Expansion to other industries
- Collaboration with Government Agencies
- Implementation of Smart Contracts

VIII. ACKNOWLEDGEMENT

We extend our sincere appreciation to our project mentor and professor, Dr. Jothi R., Professor at the School of Computer Science Engineering at the Vellore Institute of Technology in Chennai. We are deeply grateful for his unwavering support and insightful guidance during the course of our project.

We would also like to express our gratitude to the Head of Department, the Dean of Academics, and the University Dean for their encouragement and imparting their wisdom to us. We are thankful for the opportunity to take this course and work on this project.

Lastly, we would like to thank our loved ones for their constant support and for being there for us during the project. Their unwavering encouragement and support meant a great deal to us.